\documentclass[twocolumn,showpacs,amsfonts,aps,prl,floatfix,%
superscriptaddress]{revtex4} 

\usepackage{amsmath}
\usepackage{bm}
\usepackage{graphicx}

\newcommand{\fm}{\mbox{fm}}

\newcommand{\be}[1]{\begin{equation}\label{#1}}
\newcommand{\ee}{\end{equation}}

\newcommand{\vtwo}{$v_2$}
\newcommand{\eq}{{\,=\,}}

\newcommand{\rperp}{\bm{r}_\perp}

\newcommand{\Fig}[1]{Figure \ref{#1}}

\begin{document}


\title{Anisotropic flow and jet quenching in ultra-relativistic U+U collisions}
\date{\today}

\author{Ulrich Heinz}
\affiliation{Department of Physics, The Ohio State University, 
  Columbus, OH 43210, USA}
\author{Anthony Kuhlman}
\affiliation{Department of Physics, The Ohio State University, 
  Columbus, OH 43210, USA}

\begin{abstract}
Full-overlap U+U collisions provide significantly larger initial
energy densities at comparable spatial deformation, and significantly
larger deformation and volume at comparable energy density, than 
semicentral Au+Au collisions. We show quantitatively that this provides 
a long lever arm for studying the hydrodynamic behavior of elliptic 
flow in much larger and denser collision systems and the predicted 
non-linear path-length dependence of radiative parton energy loss.
\end{abstract}

\pacs{25.75.-q, 25.75.Nq, 12.38.Mh, 12.38.Qk}

\maketitle

Two major discoveries made in relativistic Au+Au collisions at the
Relativistic Heavy Ion Collider (RHIC) are the large magnitude and 
almost ideal fluid dynamical behaviour of the elliptic flow, and 
the observation of a strong suppression of high-$p_T$ particle production 
and jet quenching \cite{QM,QGP3}. Combined with a range of other
observations, they led several authors to claim creation of a 
thermalized quark-gluon plasma (QGP) in these
collisions \cite{Panic02,Gyul,Shuryak:2004cy,BM04}. While this question
is still being extensively discussed by the wider RHIC community, it is 
clear that, even if the answer is positive, we have only just scratched 
the surface of understanding the properties of this QGP.

In this Letter we address two important open questions and suggest 
that experiments studying full-overlap U+U collisions can contribute 
decisively to answering them, by exploiting the unique differences 
between the side-on-side and edge-on-edge configurations when colliding 
significantly deformed nuclei. The first concerns the observed almost 
ideal fluid dynamical behaviour of the ``elliptic flow'', described 
by the second Fourier coefficient $v_2$ of the azimuthal momentum 
distribution of emitted particles. As noted in \cite{O92}, the final 
momentum anisotropy $v_2$ is driven by the initial spatial 
eccentricity $\epsilon_x$ of the nuclear overlap region via 
anisotropic pressure gradients. Systematic studies of $v_2$ at 
midrapidity in Au+Au and Pb+Pb collisions of varying centrality at 
the Alternating Gradients Synchrotron (AGS), Super Proton Synchrotron 
(SPS), and RHIC \cite{STARv2,NA49v2} show that the ratio 
$v_2/\epsilon_x$ scales 
with the charged multiplicity density per unit transverse area, 
$\frac{1}{S}\frac{dN_{\rm ch}}{dy}$, which is proportional to the 
initial entropy density $s_0$ of the reaction zone. A similar 
scaling is seen when studying $v_2/\epsilon_x$ as a function of rapidity
in minimum bias Au+Au collisions at RHIC \cite{PHOBOSv2,Hirano01,HK04}.   
Predictions from ideal fluid dynamics \cite{KSH00} agree with the data 
only at the top RHIC energy, in almost central Au+Au collisions, and
at midrapidity, where the highest initial entropy 
densities are created. As one moves to more peripheral collisions,
lower collision energies, or away from midrapidity, the measured
elliptic flow begins to increasingly fall below the ideal fluid 
limit. According to Figure~25 in \cite{NA49v2}, the data do not seem 
to approach the ideal fluid limit gradually, but follow a
trend which seems to {\em cross} the hydrodynamic curve 
near $\frac{1}{S}\frac{dN_{\rm ch}}{dy}\approx 25/{\rm fm}^2$
\cite{fn0}. This
is unexpected since the ideal fluid value for $v_2/\epsilon_x$
is an upper limit which should not be exceeded \cite{HK02}. It is
therefore very important to check that at larger values of
$\frac{1}{S}\frac{dN_{\rm ch}}{dy}$ the data indeed settle down on 
the hydrodynamic prediction. If they do not, this would imply a stiffer
QGP equation of state than so far assumed since the hydrodynamic
value of $v_2/\epsilon_x$ increases with the speed of sound \cite{O92}.
With Au+Au collisions the only possibility to further 
raise $\frac{1}{S}\frac{dN_{\rm ch}}{dy}$ is to increase the 
collision energy, which requires waiting for the Large Hadron Collider
(LHC). We show here that with full-overlap U+U collisions at top 
RHIC energy in edge-on-edge geometry, one can increase 
$\frac{1}{S}\frac{dN_{\rm ch}}{dy}\sim 
s_0$ by ${\approx\,}55\%$, to values around 40/fm$^2$. 
This is a larger gain than between Au+Au collisions at impact 
parameters $b\eq0$ and $b{\,\simeq\,}10$\,fm.

The second major open question is which mechanism is responsible
for the large observed energy loss of fast partons travelling through
the dense medium. Fits of the energy loss data at RHIC with a theory 
based on non-Abelian radiative energy loss in a thermalized, color 
deconfined medium \cite{VG02} work well and yield initial energy 
densities consistent with those required for a successful hydrodynamic 
description of the elliptic flow \cite{Gyul}. However, the theory
makes the specific prediction \cite{BDMPS} that the dependence of the 
energy loss on the path-length through the dense medium should be 
non-linear, and this has not yet been tested. Experiments show 
that in semiperipheral Au+Au collisions fast partons travelling through 
the medium in the direction perpendicular to the reaction plane lose more 
energy than partons passing through it in the shorter in-plane direction
\cite{STARjets}. While this does prove path-length dependence of 
the energy loss, existing analyses cannot 
distinguish between different types of path-length dependences. The 
rather small size of the fireball created in semiperipheral Au+Au 
collisions does not provide much of a path-length difference between 
the in-plane and out-of-plane directions, thus limiting the resolving 
power. We show here that this situation improves dramatically 
in full-overlap U+U collisions whose initial overlap zone in the 
side-on-side configuration is about twice as large as that created in 
semiperipheral Au+Au collisions of similar eccentricity, thereby 
increasing by more than 100\% both the absolute value of the radiative 
energy loss and its difference between in-plane and out-of-plane 
directions.

Uranium-uranium collisions have been proposed before \cite{Li00,Sh00,KSH00}.
The present work
goes beyond these studies by providing quantitative calculations of
the distributions of multiplicity and spatial eccentricity in 
full-overlap U+U collisions and by presenting semiquantitative 
estimates of the energy loss of fast partons as a function of their
azimuthal emission angle. Our calculations demonstrate conclusively 
that a meaningful research program with full-overlap U+U collisions 
is experimentally feasible, and that it provides a strong lever arm 
for studying the hydrodynamic behaviour of anisotropic flow 
and the non-linear path-length dependence of non-Abelian radiative 
parton energy loss as predicted by QCD. 

We compute the initial entropy production in the $z\eq0$ transverse 
plane with a Glauber ansatz \cite{Kolb:2001qz} where a fraction 
$\alpha$ is taken to scale with the tranverse density $n_{\rm w}(\rperp)$ 
of wounded nucleons while the remainder scales with the density of binary
collisions $n_{\rm b}(\rperp)$:
\be{equation:entropy_density}
   s(\rperp;\Phi) = \kappa_s \bigl[\alpha\, n_{\rm w}(\rperp;\Phi) 
                    + (1{-}\alpha)\, n_{\rm b}(\rperp;\Phi)\bigr].
\ee
We consider only $b\eq0$ collisions with full nuclear overlap, by 
using the two forward and backward zero degree calorimeters to select 
high-multiplicity events with essentially no spectator nucleons along
the beam directions. The collision configuration is then completely 
controlled by the polar angle $\Phi$ between the beam direction and 
the symmetry axis 
of one of the two deformed U nuclei since full overlap requires the 
uranium axes to be co\-pla\-nar while their angles with the 
beam axis satisfy $\Phi_1\eq\pm\Phi_2$ (see illustration in 
Fig.~\ref{figure:mult_distribution} below) \cite{fn1}.  

We use a Woods-Saxon form for the uranium density, with 
$R(\theta)\eq(6.8\,{\rm fm})(0.91+0.26\cos^2\theta)$ for the 
nuclear radius as a function of the polar angle $\theta$ relative
to the nuclear symmetry axis and with surface thickness parameter
$\xi\eq0.54$\,fm. This gives $R_\parallel\eq7.94$\,fm and 
$R_\perp\eq6.17$\,fm, with a ratio $R_\parallel/R_\perp\eq1.29$
\cite{KSH00,Sh00,BM}; it ignores the hexadecupole moment of the U 
nucleus \cite{BM}. The normalization $\kappa_s$ in 
Eq.~(\ref{equation:entropy_density}) is adjusted to obtain a central
entropy density in $b\eq0$ Au+Au collisions of $s_0\eq117\,\fm^{-3}$
at proper time $\tau_0\eq0.6$\,fm/$c$; after hydrodynamic evolution
\cite{KSH00,Kolb:2001qz} this reproduces the charged particle
multiplicity measured at midrapidity in such collisions at  
$\sqrt{s_{NN}}\eq200$\,GeV \cite{Back:2002uc}. $\alpha$
in Eq.~(\ref{equation:entropy_density}) is fitted to the 
centrality dependence of the charged particle multiplicity in Au+Au
collisions at $\sqrt{s_{NN}}\eq200$\,GeV \cite{Back:2002uc}, by 
assuming that particle production is proportional to the total 
entropy in the transverse plane: 
$\frac{dN_{ch}}{d\eta}(b) \propto \int d^2r_\perp\ s(\rperp;b).$
The resulting fit parameter $\alpha = 0.75$ is consistent with 
results obtained in Ref.~\cite{Kharzeev:2000ph} using a different 
form of parametrization.

After fitting the Glauber model parameters to Au+Au data at 
200\,$A$\,GeV, we can predict the particle multiplicities for U+U
collisions at the same energy. Due to the binary collision
component, particle production in full-overlap U+U collisions
varies by almost 15\% between the side-on-side and edge-on-edge
configurations (see Fig.~\ref{figure:mult_distribution} below),
even though the number of wounded nucleons is almost constant \cite{fn2}.

%
\begin{figure}[hbt]
\includegraphics[width = \linewidth,clip]{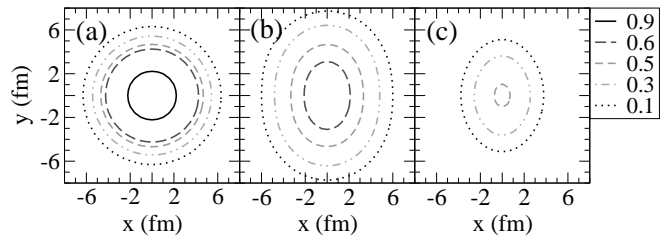}
\caption{Entropy density contours for full-overlap U+U collisions
         with $\Phi\eq0$ (a) and $\Phi\eq\frac{\pi}{2}$ (b),
         and for $b\eq7$\,fm Au+Au collisions (c). Lines show 
         specified fractions of the peak entropy density
         $s_0^{\rm UU}\eq167$\,fm$^{-3}$ in $\Phi\eq0$ U+U collisions.
}
\label{figure:entropy_contours}
\end{figure}
%
Figure~\ref{figure:entropy_contours} presents contour plots of the initial
entropy distribution in the transverse plane. The left two panels show
the profiles expected for central U+U collisions with $\Phi\eq0$ and 
$\Phi\eq\frac{\pi}{2}$, respectively. The side-on-side configuration
produces a substantial out-of-plane deformation, whereas the 
edge-on-edge configuration has a much higher peak entropy density,
due to the larger binary collision component and the smaller 
transverse overlap area. The initial eccentricity of the reaction 
zone
\be{equation:eccentricity}
  \epsilon_x(\Phi) = \frac{\int d^2r_\perp\, (y^2 - x^2)\, s(\rperp;\Phi)}
                          {\int d^2r_\perp\, (y^2 + x^2)\, s(\rperp;\Phi)}
\ee
ranges from $\epsilon_x\eq0$ in the edge-on-edge configuration to 
$\epsilon_x\eq0.25$ in the side-on-side case. The latter value is 
almost as large as for $b\eq7$\,fm Au+Au collisions 
(Fig.~\ref{figure:entropy_contours}c), 
but in this case the overlap region covers less than half the area, 
and the peak entropy density is about 25\% smaller. Averaging the 
full-overlap U+U collisions over all angles $\Phi$ gives 
$\langle \epsilon_x \rangle{\,\approx\,}0.13$, showing that the average 
reaction zone retains more than 50\% of its maximum deformation. 

\begin{figure}[h]
\includegraphics[width=0.95\linewidth,clip]{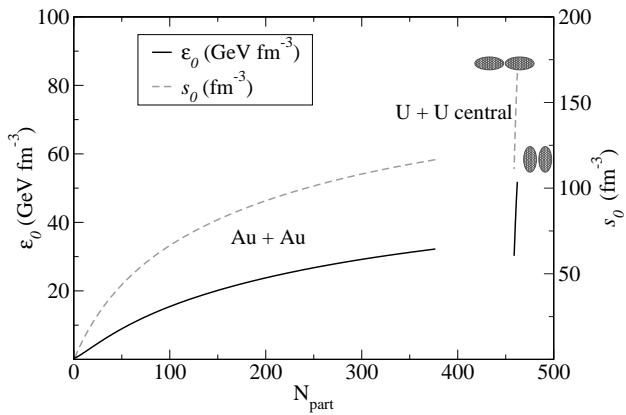}
\caption{Peak energy density (left ordinate) and entropy density 
(right ordinate) as a function of the number of participants, for
Au+Au collisions of varying impact parameter and full-overlap U+U collisions.
}
\label{figure:energy_entropy}
\end{figure}

\Fig{figure:energy_entropy} compares the peak energy and entropy 
densities in Au+Au and central U+U collisions. All curves
refer to time $\tau_0\eq0.6$\,fm/$c$ and $\sqrt{s_{NN}}\eq200$\,GeV. 
The conversion of entropy to energy density
assumes an ideal quark-gluon gas equation of state. The maximum peak 
energy density in central U+U is seen to be about 62\% larger than that 
in the most central Au+Au collisions. This gives a large lever arm to 
probe the approach to ideal hydrodynamic behavior of \vtwo.

Figure~\ref{figure:mult_distribution} shows the charged multiplicity 
distribution of full-overlap U+U collisions. To compute it we introduce 
event-by-event multiplicity fluctuations at fixed angle $\Phi$
via the probability density \cite{Kharzeev:2000ph}
\be{equation:mult_probability}
  \frac{dP}{dn\ d\Phi} = A
  \exp\left\{-\frac{(n-\bar{n}(\Phi))^2}{2a\bar{n}(\Phi)}\right\},
\ee
with $a{\,=\,}0.6$ \cite{Kharzeev:2000ph}, and integrate over
$\Phi$. Here $n$ is shorthand for $\frac{dN_{\rm ch}}{dy}$. The 
average multiplicity $\bar{n}(\Phi)$ is computed from the tranverse 
integral over Eq.~(\ref{equation:entropy_density}), using the appropriate
proportionality constant. The resulting multiplicity distribution 
in Fig.~\ref{figure:mult_distribution} exhibits Jacobian peaks near 
$\Phi\eq0$ and $\Phi\eq\frac{\pi}{2}$ since $d\bar{n}/d\Phi$ is 
smaller for angles near $0^\circ$ and $90^\circ$ than for intermediate
ones. Of course, Fig.~\ref{figure:mult_distribution} represents the 
ideal case of strictly rejecting spectators on either side of the 
collision point \cite{fn1}.
%
\begin{figure}[hbt]
\includegraphics[width=0.95\linewidth,clip]{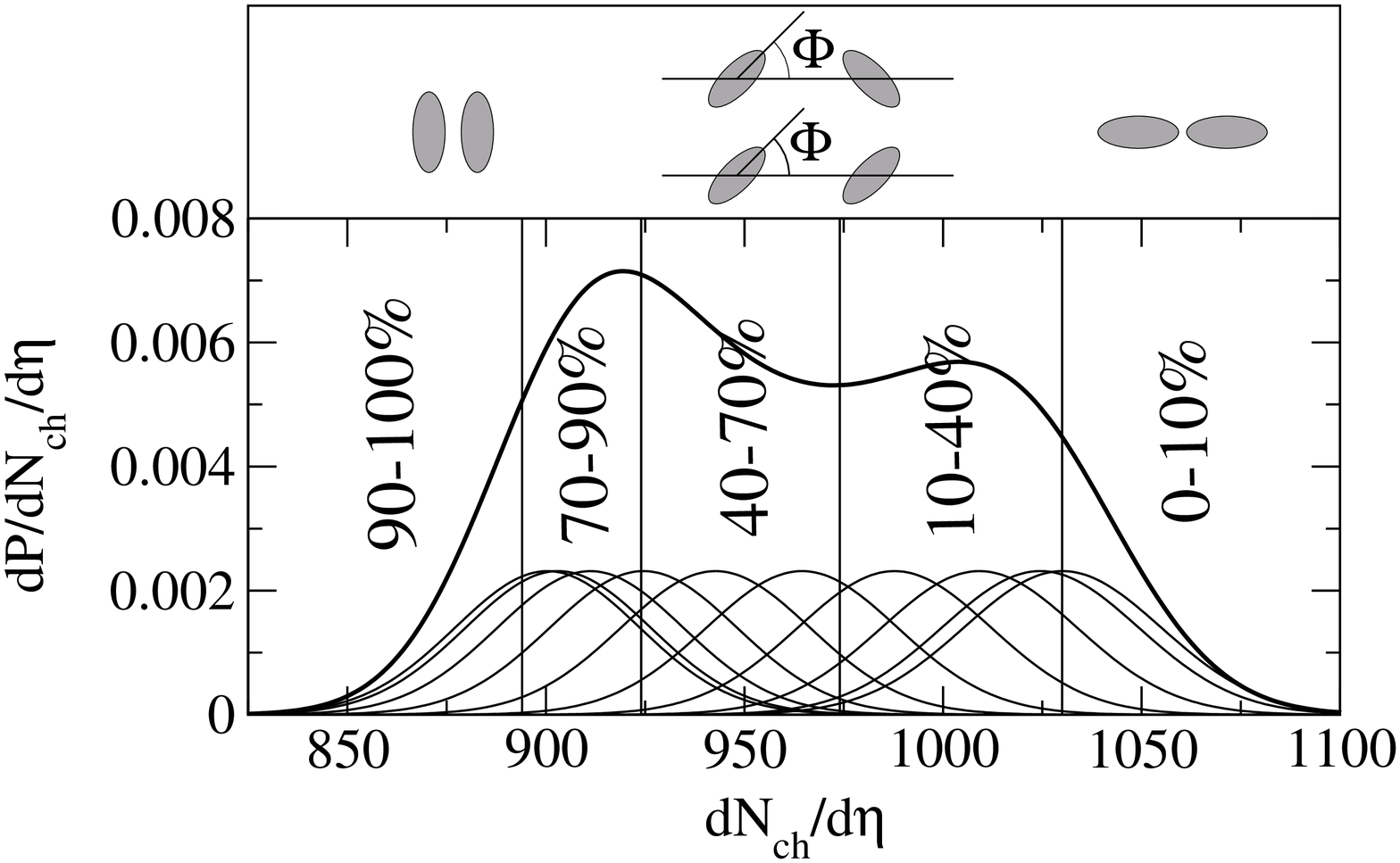}
\caption{Multiplicity distribution for full-overlap U+U
  collisions. The Gaussian curves at the bottom show individual 
  distributions for fixed angles $\Phi$ from $0^\circ$ (right) to 
  $90^\circ$ (left), in $10^\circ$ increments. The vertical
  lines cut the area under the curve according to the listed percentages.
}
\label{figure:mult_distribution}
\end{figure}
%

\begin{figure}[hbt]
\includegraphics[width=0.95\linewidth,clip]{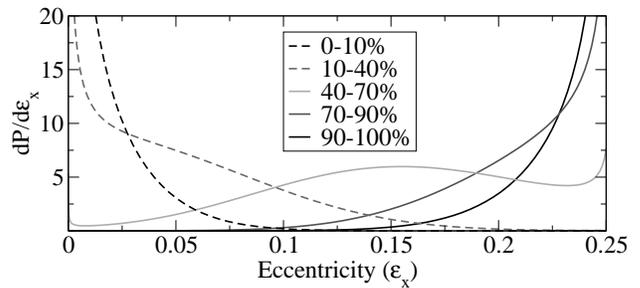}
\caption{Eccentricity distributions corresponding to the 
  multiplicity cuts shown in \Fig{figure:mult_distribution}.
}
\label{figure:ecc_distributions}
\end{figure}

This multiplicity distribution can be converted to a distribution
of eccentricities via
\be{equation:eccentricity_distribution}
  \left.\frac{dP}{d\epsilon_x} \right|_{n_0}^{n_1}
  = \frac{B}{d\epsilon_x/d\Phi}
  \int_{n_0}^{n_1} dn\
  \exp\left\{-\frac{(n-\bar{n}(\Phi))^2}{2a\bar{n}(\Phi)}\right\}.
\ee
Distributions corresponding to the cuts shown in
\Fig{figure:mult_distribution}\ are plotted in 
\Fig{figure:ecc_distributions}. One sees that by cutting on
multiplicity one can effectively select the initial eccentricity,
especially near the upper and lower end of the multiplicity
distribution.
%
%

We close by estimating the radiative energy loss, $\Delta E$, of a 
fast parton moving through the fireball medium. Our goal is not an
accurate calculation of this quantity, which would require a more
sophisticated treatment, but a qualitative comparison of the additional
reach provided by central U+U collisions compared to Au+Au.
Following Ref.~\cite{QGP3}, we therefore consider the figure of
merit
\be{equation:energy_loss}
  t \equiv \int_{\tau_0}^\infty d\tau\ \rho(\rperp(\tau),\tau)(\tau - \tau_0)
\ee
as a measure expected to be roughly proportional to 
the energy loss $\Delta E$. $\rperp(\tau)$ denotes the parton 
trajectory, and $\rho(\rperp(\tau),\tau)$ is the total parton 
density in the medium.

%
\begin{figure}
\includegraphics[width=0.95\linewidth,clip]{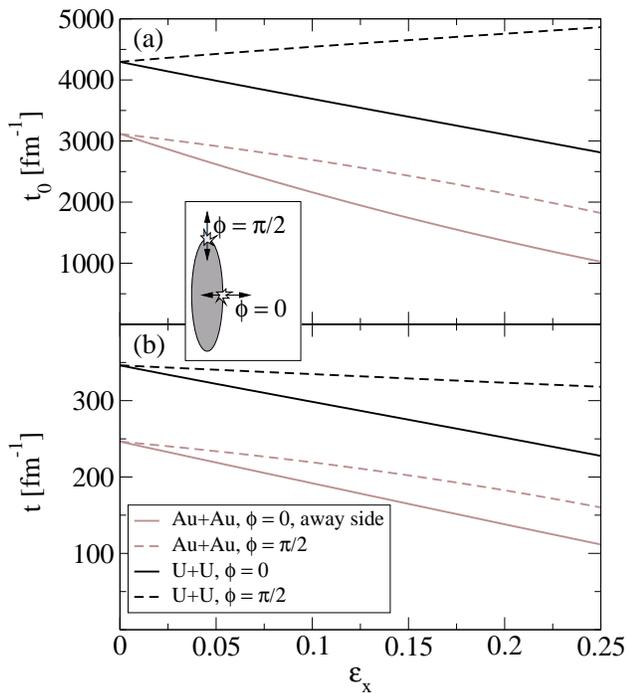}
\caption{Energy loss as a function of eccentricity for U+U (black) 
  and Au+Au collisions (gray), assuming constant density (a)
  and dilution via longitudinal expansion (b), respectively. 
  See text for discussion.
}
\label{figure:energy_loss}
\end{figure}
%

Figure~\ref{figure:energy_loss} compares the energy loss for inward-moving
partons produced near the edge \cite{fn3} of the fireball (see inset)
for Au+Au and full-overlap U+U collisions. The top panel (labelled $t_0$) 
assumes that the parton density does not change while the parton passes 
through the fireball; the bottom panel accounts for 
dilution of the density by longitudinal expansion using
$\rho(\rperp,\tau)\eq\frac{\tau_0}{\tau}\rho(\rperp,\tau_0)$.
The energy loss is calculated as a function of source eccentricity
$\epsilon_x$, and we compare it for partons emitted into ($\phi\eq0$) 
and perpendicular to the reaction plane ($\phi\eq\frac{\pi}{2}$)
\cite{fn4}. In Au+Au collisions, changing the eccentricity in order
to study the path-length dependence of energy loss requires going to
more peripheral collisions which produce smaller fireballs. In 
full-overlap U+U collisions, the eccentricity can be increased 
without decreasing the fireball size (although the density decreases
somewhat). Were it not for the dilution of the density due to 
longitudinal expansion, this would in fact lead to {\em larger} energy
loss for out-of-plane partons emitted from side-on-side collisions 
compared to edge-on-edge collisions (dashed line in 
Fig.~\ref{figure:energy_loss}a). When longitudinal expansion is 
included (bottom panel), the energy loss for out-of-plane emitted 
partons from U+U collisions is still almost independent of eccentricity, 
whereas the in-plane energy loss decreases by about 35\% between 
edge-on-edge and side-on-side collisions. In Au+Au collisions the 
energy loss decreases with increasing $\epsilon_x$ in both cases 
(by 35\% and 55\%, respectively), leading to an overall loss of 
discriminating power on its path-length dependence. For the largest
eccentricities, the {\em difference} in energy loss between out-of-plane
and in-plane emission in U+U is more than twice that which 
can be achieved in Au+Au collisions. The {\em total} energy loss is also
larger by up to a factor 2, causing significant jet quenching and 
reduced particle production up to much larger $p_T$ values
in full-overlap U+U collisions than in Au+Au collisions.

We thank R.~Furnstahl and I.~Tornes for helpful technical comments
and acknowledge support by the U.S. DOE under contract DE-FG02-01ER41190.



\begin{thebibliography}{99}
\bibitem{fn1}
Here we do not attempt to estimate the smearing effects from  
collisions with small nonzero impact parameters or with symmetry 
axes that are not exactly coplanar. While they can be 
minimized by imposing stringent experimental cuts, they cannot be
completely avoided. Their study requires a multidimensional 
simulation in the space defined by the impact parameter $b$ and 
the two Euler angles associated with the relative orientation of 
the U nuclei; we postpone this for a longer publication.
\bibitem{fn2}
For side-on-side collisions the fraction of surface nucleons 
passing through without scattering is slightly larger.
\bibitem{fn3}
The ``edge'' is defined as the parton density contour corresponding to 
10\% of the peak value.
\bibitem{fn4}
We only consider partons created at the specific points shown in the inset
of Fig.~\ref{figure:energy_loss}. A more quantitative study will
require averaging over all possible production points.
\end{thebibliography}

\end{document}